%Paper: 9204005
%From: creutz@wind.phy.bnl.gov (Michael Creutz)
%Date: Thu, 16 Apr 92 09:31:37 EDT

\input mtexsis
% this uses the texsis macro package
\preprint
\superrefsfalse
\titlepage
\title
Microcanonical Cluster Monte Carlo
\endtitle
\author
Michael Creutz
Physics Department
Brookhaven National Laboratory
Upton, NY 11973
\medskip
creutz@wind.phy.bnl.gov
\endauthor
\abstract
I propose a numerical simulation algorithm for statistical systems
which combines a microcanonical transfer of energy with global
changes in clusters of spins.  The advantages of the
cluster approach near a critical point augment the speed increases
associated with multi-spin coding in the microcanonical approach.
The method also provides a limited ability to tune the average
cluster size.
  \endabstract
\disclaimer{DE-AC02-76CH00016} \endtitlepage

Monte Carlo simulation is now a major tool for the study of both
critical phenomena in condensed matter physics and non-perturbative
field theory in particle physics.  With continual tuning
over the years, current algorithms are well adapted to today's
supercomputers.  Nevertheless, on large systems near a critical
point, severe slowing of the evolution to independent states
encourages the search for yet better methods.  Two intriguing and
apparently unrelated ideas for such improvement are cluster
algorithms and microcanonical techniques.

Swendsen and Wang \reference{SW} R.H.~Swendsen and J.-S.~Wang
\PRL \vol{58}, 86 (1987)\endreference
proposed cluster algorithms as a way to make coherent long range
changes in a single Monte Carlo updating step.  They were motivated
by the rather slow probing of important long distance physics by
local algorithms when the couplings of a system are near a critical
point. Later Wolff \reference{Wolff}U.~Wolff, \PRL \vol{62}, 361
(1989)\endreference  presented a powerful variation on the approach,
where at each step a single large cluster of variables is formed and
modified.

Another useful scheme is the microcanonical
Monte Carlo approach\reference{demons}M.~Creutz, \PRL \vol{50},
1411 (1983)\endreference
.  Here a
set of additional variables, referred to as demons, are introduced to
transfer energy around the system.  During the updating, the combined
energy of the system of interest and the auxiliary variables is held
absolutely constant.  In this approach quantities such as the
temperature are outputs of the simulation, being extracted from the
distribution of demon energies.  The primary advantage is that the
demons can take a very simple form, and for discrete systems they can
be treated entirely with simple bit manipulation.  With many demons
stored in a few computer words, an effective parallelization is
possible on conventional serial computers.  In simple cases no
floating point arithmetic is ever needed, and many individual spins
can be updated in parallel via bitwise operations.  This gives
programs for the Ising model which run about an order of magnitude
faster than conventional approaches. \reference{fast ising}M.~Creutz,
G.~Bhanot, and H.~Neuberger, Nuclear Physics \vol{B235[FS11]}, 417
(1984); M.~Creutz, P.~Mitra, and K.J.M.~Moriarty, Comp. Phys. Comm.
\vol{33}, 361 (1984); M.~Creutz, P.~Mitra, and K.J.M.~Moriarty, J.
Stat. Physics \vol{42}, 823 (1986); M.~Creutz and K.J.M.~Moriarty,
Comp. Phys. Comm. \vol{39}, 173 (1986); M.~Creutz, K.J.M.~Moriarty
and M.~O'Brien, Comp. Phys. Comm. \vol{42}, 191 (1986)\endreference

In this paper I present a combination of these ideas, a cluster
algorithm where the cluster growth is determined entirely by bit
manipulations involving a set of microcanonical demons.   The primary
gain is increased simulation speed while retaining the
advantages of a global updating scheme near a critical point.  In
addition, discussions of detailed balance are particularly
straightforward in the microcanonical language, simplifying
justification of variations in the cluster algorithms.  The
microcanonical approach also introduces new parameters for tuning the
cluster size.  As with the local microcanonical method, the
temperature is determined as a function of the constant total system
energy.   Wolff \reference{wolff2} U.~Wolfe, \NP \vol{ B300[FS22]},
517 (1988)\endreference
  has considered another hybrid of the cluster and microcanonical
approach, where, rather than the energy, the total area of the
cluster boundaries was held fixed.  This allowed him to directly and
efficiently study the diluted Ising systems of ref.~\reference{FK}
P.W. Kasteleyn and C.M. Fortuin, J. Phys. Soc. Japan \vol{26}
(Suppl.), 11 (1969); C.M. Fortuin and P.W. Kasteleyn, Physica
\vol{57}, 536 (1972); C.M. Fortuin, Physica \vol{58}, 393 (1972);
C.M. Fortuin, Physica \vol{59}, 545 (1972)\endreference.

For simplicity I begin with the simple Ising model.  On
each site $i$ of an arbitrary lattice lies a spin variable $\sigma_i$
taking values in $\{1,-1\}$.  The energy of each bond is lowest if
the neighboring spins have the same value, and is increased by two
units for antiparallel spins.  Thus I consider the Hamiltonian
$$
H_\sigma=-\sum_{\{i,j\}} \sigma_i \sigma_j
\eqno(1)
$$
where $\{i,j\}$ denotes the set of nearest neighbor pairs, each such
pair appearing once in the sum.  I am interested in the statistical
mechanics of this system at inverse temperature $\beta$, and thus
consider the partition function
  $$
Z_\sigma=\sum_{\{\sigma\}} \exp(-\beta H_\sigma).
\eqno(2)
$$
As is well known \reference{ising} L.~Onsager, \PR \vol{65}, 117
(1944); T.D.~Schultz, D.C.~Mattis, and E.H.~Lieb,
Rev.~Mod.~Phys.~\vol{36}, 856 (1964)\endreference
this model has a second order magnetic phase transition at
$\beta=(1/2)\log(1+\sqrt 2)=0.44068...$.

As in ref.\ref{demons} I now augment this system with a set of
auxiliary variables called "demons."  In that reference the demons
were associated with the lattice sites.  Here, however, I place them
on the system bonds.  Each of these demons carries a sack of energy
which it can use to "flip" or change the state of the
bond it occupies.  Thus for each neighboring pair of sites $i$ and $j$, I
associate a demon energy $D_{ij}\equiv D_{ji} \ge 0$.  For the Ising
case the bond energies always change in steps of two, so I am free
to restrict the demon energies to be non-negative even integers.  In
this case it is also convenient to place a capacity limit on the
demon in the form of an upper bound for its energy $D_{ij}\le
D_{max}$.  This allows storage of the demon energies in a few bits.
For example, with two bits per demon I can consider the individual
demon energies to lie in the set $\{0,2,4,6\}$, and store $32$ such
demons in two $32$ bit computer words.  As with the local
microcanonical approach, the upper limit on $D_{ij}$ still leaves the
algorithm exact, although it will modify the cluster shapes.

The total energy for the coupled system is simply
  $$
H=-\sum_{\{i,j\}} \sigma_i \sigma_j+D_{ij}. \eqno(3)
 $$
The corresponding canonical partition function is
 $$
Z=\sum_{\{\sigma,D\}}
\exp(-\beta H). \eqno(4)
$$
This immediately factorizes into contributions from the lattice and
the demons.  It is only through a microcanonical constraint that
these variables become coupled.

The algorithm consists of three parts, cluster growing, cluster
flipping, and demon shuffling.  For the demon shuffling one merely
moves the demons around on the bonds to new locations.  This can be
done in an arbitrary way, as can easily be seen from the fact that
the Hamiltonian in Eq.~(3) leaves the demons uncoupled.  Indeed, the
separate bits of the demons are also uncoupled; thus, the first bits
could be shuffled separately from the second bits.  (For the test
simulations below, the bits were shuffled together.)

For the cluster growing, I divide the demons into two sets,
"contented" and "frustrated."  A demon is contented if it posesses an
amount of energy which allows a change in the state of the
currently occupied bond.  Thus, if the neighboring spins are parallel
and the demon has 2 or more units of energy, then it is happy.  If
the demon cannot accomodate the change in the bond energy, it is
frustrated.  This might occur for parallel spins if the demon has no
energy, or for antiparallel spins if the demons energy supply is full
and cannot accept more. A cluster is now defined as a complete set of
sites joined by frustrated demons.  All bonds on the exterior
boundary of such a cluster carry contented demons.  The contentedness
of the demon on any given bond corresponds closely with the bond
occupation variables used in ref.~\ref{FK}.

I now come to the cluster flipping stage.  Here I could either follow
the approach of Swendsen and Wang \ref{SW} or the variation of Wolff
\ref{Wolff}. In the former case, the lattice is divided into clusters
as above, and with a random probability all spins in each cluster are
either flipped or not.  In the Wolff approach, a single random site
is chosen, and the corresponding cluster has all of its spins
inverted.  This picks a given cluster with a probability proportional
to its volume, and thus gives larger average cluster sizes.  For the
remainder of this discussion I consider this single cluster approach.
When the cluster(s) of spins is(are) flipped, the appropriate changes
of the demon energies on the cluster edges must also be made.  Thus
the energy in Eq.~(3) is absolutely conserved.

The justification of the procedure is a particularly simple
application of detailed balance.  After a cluster is  flipped, all
demons retain their contented or frustrated state.  Thus if we were
to regrow the clusters, their shapes would be unchanged.  If the
demons are not moved and a cluster is grown from the same point, a
second application of the algorithm will return exactly to the
starting state.  With an ensemble in  equilibrium, all states of
equal energy are equally likely.  Under the algorithm the states of
the coupled demon-spin system break up into pairs of degenerate
states.  These states just flip back and forth between each other,
and remain equally likely.  The demon shuffling stage then changes
the cluster breakup for later steps.

As discussed in ref.~\ref{demons}, the microcanonical approach has no
explicit coupling parameter in the algorithm.  In equilibrium, the
temperature is an output and depends on the total system energy.
I can easily find it from the distribution of demon
energies.  Here the separate demon bits are uncoupled; so, I can
extract the inverse temperature from the expectation of a
single demon bit.  For example, I have
  $$
\beta=-(1/2) \log(-1+1/P) \eqno(4)
$$
where $P$ is the average fraction of the demons with value 2 or
6.  This probability is quickly found by counting the number of
times the lowest demon bit is set.

The primary advantage of the approach is that all necessary
arithmetic can be done entirely by bitwise operations.  For example,
on a thirty two bit machine a group of 32 adjacent spins can be
stored in a single computer word $I$, and 32 two bit demons can be
stored in two computer words $D1$ and $D2$.  Using shifts and logical
operations, the calculations to find the frustrated and
contented demons and to grow the cluster can be done for all 32
demons simultaneously.  I note in passing that, unlike in the local
approaches, no division of the lattice into independent sets of
sites, such as a checkerboard, is involved.

The use of bit manipulation gives a potential speed increase over the
conventional cluster approach, wherein a bond is included in a
cluster based on a floating point comparison with a random number.
On the other hand, the gain is not as much as for  local algorithms
because much of the computer time is spent growing the cluster, and
once the frustrated bonds are found  this is essentially the same in
both algorithms.  The cluster growth can still be done by bit
manipulation, but with a porous cluster much of the work involves
irrelevant sites.

Wolff \ref{Wolff} observed that for the Ising model his average
cluster size is tied to the magnetic susceptibility.  As our demons
have upper as well as lower bounds on their energies, this connection
is less precise here, with a bond connecting antiparallel spins
having an additional chance to be included in the cluster.  In
general, the clusters tend to be small at high temperatures where it
is easy to satisfy the demons desires, while they become increasingly
dense at low temperatures where only a few demons will have the
required energy to excite their bonds.  In Fig.~(1) I show the
average cluster size as a function of beta for demons carrying from
one to four bits.   Note how the average cluster size
increases as the number of demon bits is reduced.    This dependence
is essentially invisible with 3 or more bits.  Indeed, occupancy of
the higher bits is exponentially suppressed by a Boltzmann factor.

Intuitively, the most independent configuration after a single step
should be obtained when about half the spins are flipped.  In the
Ising case this occurs conveniently near the critical temperature.
At high temperatures the clusters are quite small, and the algorithm
seems to have no advantage over a conventional local approach.
Conversely, at low temperatures the clusters dominate the lattice
and primarily serve to flip the lattice back and forth between
opposite magnetizations.  Again, the advantage over a local algorithm
is limited.  I note, however, that if a hot lattice is rapidly
quenched to a low temperature state where several distinct domains
are frozen in, the cluster algorithm can easily generate a clusters
which fill a single domain.  In this case the approach is quite
efficient at relaxing the system to its true ground state.

Fig.~(1) shows that with single bit demons, for all energies the
clusters tend to be quite large, and most spins flip at each
step.  While the algorithm thus will require more iterations to reach
a truly independent configuration, it is perhaps worth noting that in
this single bit limit the algorithm is particularly simple.
Contented demons, which form the cluster edge, are those where of the
three bits consisting of the spins and the demon himself an odd
number are set.  This can be determined by a simple exclusive or
operation between the three bits.

Varying the number of demon bits gives some control over the average
cluster size.  This control could become an essentially continuous
parameter because not all demons need have the same number of bits.
For simulations of continuous fields, an upper bound on the demon
energies similarly will increase the average cluster volume. In
general additional constraints on the demon energies work to increase
the number of frustrated demons.  As it is forbidden to exclude any
frustrated demon bonds from the cluster, such constraints will always
tend to increase the average cluster size.  Unfortunately, it is
unclear how to modify the approach to obtain smaller clusters.

In general, cluster algorithms are more limited in their
applicability to systems where the spins are continuous variables.
The usual approach is to imbed a $Z_2$ symmetry in the spin
manifold\ref{Wolff}.   The microcanonical idea is easily adapted to
such systems.  In this case, however, the bit manipulation advantages
of the discrete models are lost.

To illustrate one way to imbed a $Z_2$ transformation in a more
complicated model, let me generalize to spins $\sigma_i$ taken as
elements of a group $G$.  For the spin energy associated with a bond
connecting sites $i$ and $j$ I consider ${\rm Re}\Tr(\sigma_i^\dagger
\sigma_j)$, where Tr denotes a group character, perhaps the trace
in the fundamental representation.  As before, I place auxiliary
demon variables on the lattice bonds, with the demon energies now
being positive real numbers $D_{ij}\ge 0$.

To keep the detailed balance condition simple, I again arrange
transitions between pairs of equal energy states.  For the updating
of a cluster, I take a random element $g$ of the group to flip the
cluster spins around. As the distance from $g$ to some spin
$\sigma$ is  $g\sigma^{-1}$, I perform the flip by multiplying
$\sigma$ on the left by the square of this distance, making the new
value for the spin $g\sigma^{-1}g$.  Note that this transformation
has the $Z_2$ property of returning to the original spin if repeated
twice with the same $g$.  Considering such a change on a spin in the
partially grown cluster, I determine if the demon on a bond emerging
from the cluster has enough energy to make the corresponding change.
If not, the demon is frustrated and the adjacent spin joins the cluster.

In Fig.~(2) I show the beta dependence of average cluster size for
this algorithm as applied in two dimensions to the U(1) or "X-Y"
model, where $\sigma$ is a complex number of unit magnitude.  At
high temperatures the clusters are again small, but for couplings of
order unity, where lies the primary physical interest in this model
\reference{xy} R.~Gupta, J.~DeLapp, G.G.~Batrouni, G.C.~Fox,
C.F.~Baillie, and J.~Apostolakis, \PRL \vol{61}, 1996 (1988)\endreference
, the clusters cover a fair fraction on the system and should give
an efficient decorrelation time.  For this figure no upper bound was
placed on the demon energies; so, the results should be essentially
identical to the canonical approach.

The microcanonical approach is easily modified
to a canonical method in the same way that the local
microcanonical algorithm of ref.~\ref{demons} reduces to the canonical
Metropolis {\it et al.}\reference{metro}N. Metropolis, A.W. Rosenbluth,
M.N. Rosenbluth, A.H. Teller, and E. Teller, J. Chem. Phys. \vol{21}
(1953) 1087\endreference
procedure.  For this reduction, I refresh the demons after each
cluster step; i.e. I replace the demon energies with a random
positive numbers selected with Boltzmann probablity $\exp(-\beta
D_{ij})$. As for the local case, for large volumes and when the demon
scrambling takes place over long distances, the canonical and
microcanonical evolutions become essentially indistinguishable.  For
discrete systems the extra arithmetic involved in going to the
canonical scheme may become appreciable.

In summary, I have presented a simulation scheme which combines
features of the cluster algorithms of refs.~\ref{SW}\ref{Wolff} with
the fast microcanonical approach of ref.~\ref{demons}.  This enables
simulations done entirely by bit manipulation, permitting parallel
operations on conventional computers.  Furthermore, the introduction
of the auxiliary demon variables provides a particularly simple way
to justify detailed balance for cluster algorithms in general.

\bigskip {\noindent ACKNOWLEDGEMENTS}

I thank A.~Gocksch, P.~Hsieh, Y.~Shen, and A.~Sokal for useful
discussions.

\vfill\eject
      {\noindent  REFERENCES}
      \bigskip
      \ListReferences

\vfill\eject
{\noindent Figure Captions}

Fig.~1. The average cluster volume divided by the total system volume
as a function of the inverse temperature.  The results are for the
two dimensional Ising model on a 320 by 320 lattice.  The squares,
diamonds, crosses, and plus signs are for are for one through four
bit demons, respectively.

Fig.~2. The average cluster volume divided by the total volume as a
function of the inverse temperature for the $U(1)$ spin model in
two dimensions on a 50 by 50 lattice.

\vfill\eject\bye